\documentclass[twocolumn, pra, superscriptaddress]{revtex4-1}
\usepackage{amssymb}
\usepackage{amsmath}
\usepackage{epsfig}
\usepackage{color}
\usepackage{graphics, graphicx}
\usepackage{bbold}
\usepackage{psfrag}
\usepackage{mathcomp}
\usepackage{subfigure}
\usepackage{verbatim}
\usepackage{color}
\usepackage[colorlinks, citecolor=blue]{hyperref}

\begin{document}

\title{Repulsive Fermi polarons with negative effective mass}
\author{Wenliang Li}
\affiliation{Beijing National Laboratory for Condensed Matter Physics, Institute of Physics, Chinese Academy of Sciences, Beijing 100190, China}
\affiliation{University of Chinese Academy of Sciences, Beijing 100049, China}
\author{Xiaoling Cui}
\email{xlcui@iphy.ac.cn}
\affiliation{Beijing National Laboratory for Condensed Matter Physics, Institute of Physics, Chinese Academy of Sciences, Beijing 100190, China}
\date{\today}
\begin{abstract}
Recent LENS experiment on a 3D Fermi gas has reported a negative effective mass ($m^*<0$) of Fermi polarons in the strongly repulsive regime. There naturally arise a question whether the negative $m^*$ is a precursor of the instability towards phase separation (or itinerant ferromagnetism). 
In this work, we make use of the exact solutions to study the ground state and excitation properties of repulsive Fermi polarons in 1D, which can also exhibit a negative $m^*$ in the super Tonks-Girardeau regime. By analyzing the total spin, quasi-momentum distribution and pair correlations, we conclude that the negative $m^*$ is irrelevant to the instability towards ferromagnetism or phase separation, but rather an intrinsic feature of collective excitations for fermions in the strongly repulsive regime. Surprisingly, for large and negative $m^*$, such excitation is accompanied with a spin density modulation when the majority fermions move closer to the impurity rather than being repelled far away, contrary to the picture of phase separation. These results shed light on the recent observation of negative $m^*$ in the 3D repulsive Fermi polarons. 
\end{abstract}

\maketitle

\section{Introduction}

 Whether itinerant fermions can become ferromagnetic in the presence of strong repulsion is a long-standing and challenging many-body problem. In recent years this problem has intrigued great interests in the field of ultracold atoms\cite{Ketterle1, Ketterle2, Italy, Allan, Para, Zhai, Troyer, Trivedi, Heiselberg, Conduit, Cui-Ho, Pilati, HuHui}, taking advantage of the highly tunable interactions via Feshbach resonance. Compared to an equal mixture of spin-1/2 fermions, the Fermi polaron system, which involves impurity fermions embedded in a Fermi sea of majority ones, can be a more convenient testbed for this problem. 
The idea is to examine whether a full ferromagnetic state, i.e., a Fermi sea of identical fermions, is energetically stable against a single spin flip\cite{Cui}. 
The system is now well known as the repulsive Fermi polaron\cite{Cui, Bruun, Bruun2}, which has been 
successfully explored in a number of cold atoms experiments\cite{Grimm, Kohl, Grimm2016, Italy2}. 

Recently, the LENS group has reported the observation of repulsive Fermi polarons with well-defined quasi-particle behavior in a 3D Fermi gas\cite{Italy2}. In the strongly repulsive regime, it was found that the polaron energy can exceed the Fermi energy, and more interestingly, the polaron can host a large and negative effective mass $m^*$. 
The negative $m^*$ has been interpreted as an indication of instability towards phase separation\cite{Italy2}, i.e., by acquire a finite momentum the impurities tend to depart from the majority cloud to form ferromagnetic domains. Nevertheless, such interpretation has not been experimentally verified. How to understand the negative $m^*$ and its associated instability is still an open question up to date, which is crucially important for understanding the nature of strongly repulsive fermions and for definitively identifying itinerant ferromagnetism in future experiment.

As we know, the strongly interacting fermion system in 3D is notoriously difficult to solve, and the theoretical tools are usually based on certain approximations that very often lead to unreliable results. This is particularly true for the system of repulsive polaron, since it lies in an excited upper branch of atomic system rather than the ground state. 
On the other hand, one might be able to gain important insights 
by studying the counterpart problem in 1D, where the exact solutions can be accessible\cite{Guan_review} and effective spin-chain models can be established in strong coupling limit\cite{Guan, Santos, Zinner, Pu, Levinsen, Yang-Cui, Jochim_chain}. For 1D fermion system, the ferromagnetic transition has been exactly proved with arbitrary number and arbitrary potential\cite{Cui-Ho2}, and the polaron problem in continuum has also been exactly solved by McGuire in 1960's\cite{McGuire1, McGuire2} and recently by Guan\cite{Guan_polaron}. In particular, a negative effective mass ($m^*<0$) has been demonstrated for the excited upper branch of 1D Fermi polarons with attractive coupling\cite{McGuire2}, i.e., in the fermionic super Tonks-Girardeau regime\cite{Chen_STG, Girardeau}. 
Given the same behavior of $m^*$ in 3D experiment\cite{Italy2}, it seems that the negative $m^*$ is an outcome uniquely driven by strong repulsions but irrelevant to the physical dimension of the system.

In this work, we utilize the exact solutions of 1D repulsive Fermi polaron, as presented in Refs.\cite{McGuire1, McGuire2}, to study its ground state and low-lying excitation properties, including the total spin, quasi-momentum distribution and pair correlations. We find that the excited states always have the same total spin as the ground state regardless of the sign of $m^*$, and there is no ferromagnetic component. 
This means that the negative $m^*$ is not associated with instability towards ferromagnetism or phase separation. Rather, it originates from a special type of collective excitation in the repulsive branch, which can be reflected in the changes of quasi-momentum distribution and pair correlations during the excitation.  
Surprisingly, for large and negative $m^*$, the excitation gives rise to a spin density modulation where the majority fermions move closer to the impurity rather than being repelled far away, contrary to the picture of  phase separation. 
These results shed insight on the negative $m^*$ 
as observed in the recent LENS experiment\cite{Italy2}. 

The organization of the rest of this paper is as follows. We will first illustrate the exact solutions and point out the negative effective mass in section II. Using the exact solutions, we will then analyze the ground-state and excitation properties of this system in section III. Finally we summarize the work in section V. 

\section{Exact solution of 1D Fermi polaron} 

We start by presenting the exact solutions of 1D Fermi polaron problem as derived by McGuire\cite{McGuire1, McGuire2}. Consider a spin-1/2 fermion system in which a single spin-$\downarrow$ interacts with the rest $N-1$ spin-$\uparrow$ fermions via coupling $g$, under the periodic boundary condition the polaron energy can be written as $E=\sum_{i=1}^N k_i^2/(2m)$, with $m$ the mass and $k_i\ (i=1,..N)$ the quasi-momentum ($\hbar=1$ in this paper). Define $z_i=k_iL/2$ ($L$ is the system size), $z_i$ can be obtained by finding the $N$ roots of the following equation:
\begin{equation}
az-\cot z=c, \label{eq_z}
\end{equation}
subject to a constraint due to periodic boundary condition:
\begin{equation}
\sum_{i=1}^N z_i=n\pi. \ \ \ \ (n=0,1...) \label{PBC}
\end{equation}
Here $a=4/(mgL)$, and $c$ is a constant. The ground state is associated with $c=0$ and $n=0$, and thus with total momentum $Q=0$; while the collective excited states are with $c\neq 0$, $n\ge 1$ and $Q\neq 0$. 

The associated polaron wave-function is given by: 
\begin{equation}
|\Psi\rangle = \int dx_1...dx_N \psi({\bf x}) 
     \psi^{\dagger}_{\downarrow}(1)  \psi^{\dagger}_{\uparrow} (2)...\psi^{\dagger}_{\uparrow} (N) |0\rangle,
\label{wf}  
\end{equation}
with ${\bf x}=(x_1,...x_N)$ and
\begin{eqnarray}
\psi({\bf x}) &&= \theta(x_1<x_2<...<x_N)\phi({\bf x}) +\nonumber\\
&&\sum_{i=2}^{N}\theta(x_2<... \ x_{i}<x_1<x_{i+1}...<x_N)\phi|_{x_{2...i}\rightarrow x_{2...i}+L};\nonumber\\ \label{wf_2}  
\end{eqnarray}
$\phi({\bf x})$ is a Slater Determinate:
\begin{equation}
\phi({\bf x})=\left|
\begin{array}{cccc}
\alpha_1 e^{ik_1x_1} & \alpha_2 e^{ik_2x_1} & \cdots & \alpha_2 e^{ik_Nx_1}\\
e^{ik_1x_2} & e^{ik_2x_2} & \cdots & e^{ik_Nx_2} \\
\vdots & \vdots & \vdots & \vdots \\
e^{ik_1x_N} & e^{ik_2x_N} & \cdots & e^{ik_Nx_N}
\end{array}
\right| \label{wf_3}
\end{equation}
with $\alpha_i=1-e^{ik_iL}$.

Given Eqs.(\ref{eq_z},\ref{PBC}), one can solve all $\{k_i\}$ for the ground state or excited states, and then obtain the energy $E$ and the effective mass $m^*=(\partial^2 E/\partial Q^2)^{-1}$. Since we are interested in the repulsive scattering branch with $E>0$, we will only collect the real solutions of $k_i$ for all the couplings. In this way, we will consider the super Tonks-Girardeau(sTG) regime of fermions in the attractive coupling $(g<0)$ side\cite{Chen_STG, Girardeau}.

\begin{figure}[t]
\includegraphics[width=8cm, height=10cm]{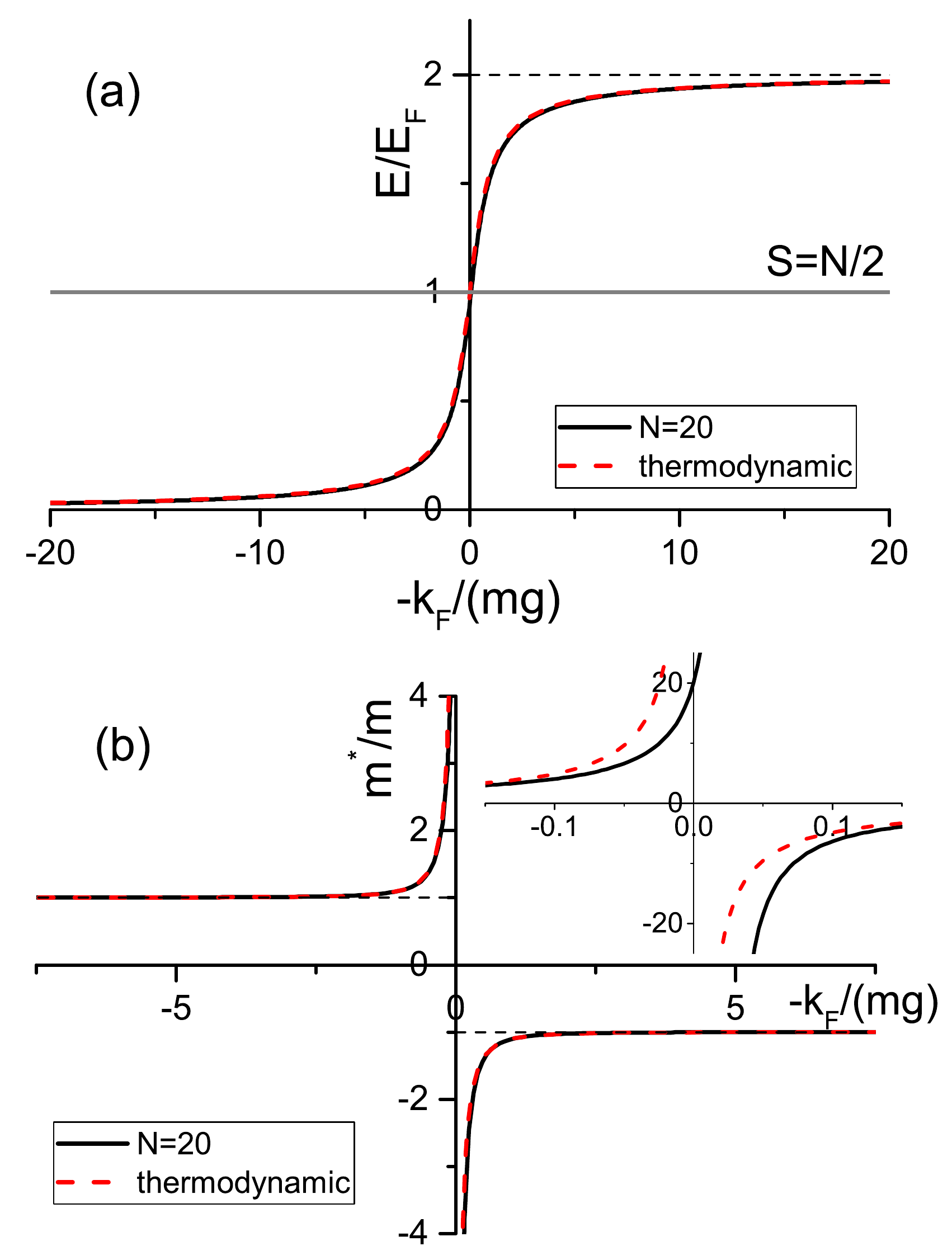}
\caption{(Color online.) Energy $E$(a) and effective mass $m^*$(b) as a function of $-1/g$ for 1D repulsive Fermi polarons with total number $N=20$. For comparison, the thermodynamic predictions\cite{McGuire1,McGuire2} are also shown (red dashed lines). 
The gray horizontal line in (a) shows the energy ($E=E_F$) of ferromagnetic state with total spin $S=N/2$. Inset of (b) shows $m^*$ in a larger range in the strong coupling regime. Here the units of momentum and energy are respectively $k_F$ and $E_F$. } \label{fig1}
\end{figure}

In Fig.\ref{fig1}, we show $E$ and $m^*$ as a function of $-1/g$ given the total number $N=20$. Here we define the Fermi momentum as $k_F=N\pi/L$, and use $k_F$ and $E_F=k_F^2/(2m)$ as the units of momentum and energy. We can see that as increasing $-1/g$ from $-\infty$(weak repulsion) to $0^-$ (hard-core or the TG limit), $E$ continuously increase from $0$ to $E_F$, and $m^*$ increase from the bare mass $m$ to $Nm$ (see inset of Fig.\ref{fig1}(b)). 
Further increase $-1/g$ into the sTG side, $E$ continues to increase, while $m^*$ undergoes a resonance at a small $-1/g$, i.e, changes from large positive to large negative. In the limit of $-1/g\rightarrow\infty$, we have $E\rightarrow 2E_F$ and $m^*\rightarrow -m$.

In Fig.\ref{fig1} we also compare the numerical results of $N=20$ system (black solid lines) with the predictions in the thermodynamic limit\cite{McGuire1,McGuire2} (red dashed lines).
We can see that the two results match quite well in most of coupling regime, except that in the thermodynamic limit $m^*$ diverges exactly at $g=\infty$ while for finite number of particles it is $Nm$ at $g=\infty$ and diverges at a finite $-k_F/(mg)\sim 0.02$ (see inset of Fig.\ref{fig1}(b)). We have checked that as increasing $N$, the numerical results gradually approach the thermodynamic predictions. 

\section{Ground state and excitation properties}

In this section, we will utilize the exact wave function in Eq.\ref{wf} to extract the essential ground-state and excitation properties of 1D repulsive polarons, which have not been revealed in previous studies including Refs.\cite{McGuire1, McGuire2}.

\subsection{Quasi-momentum distribution}

\begin{figure}[h]
\includegraphics[width=9cm]{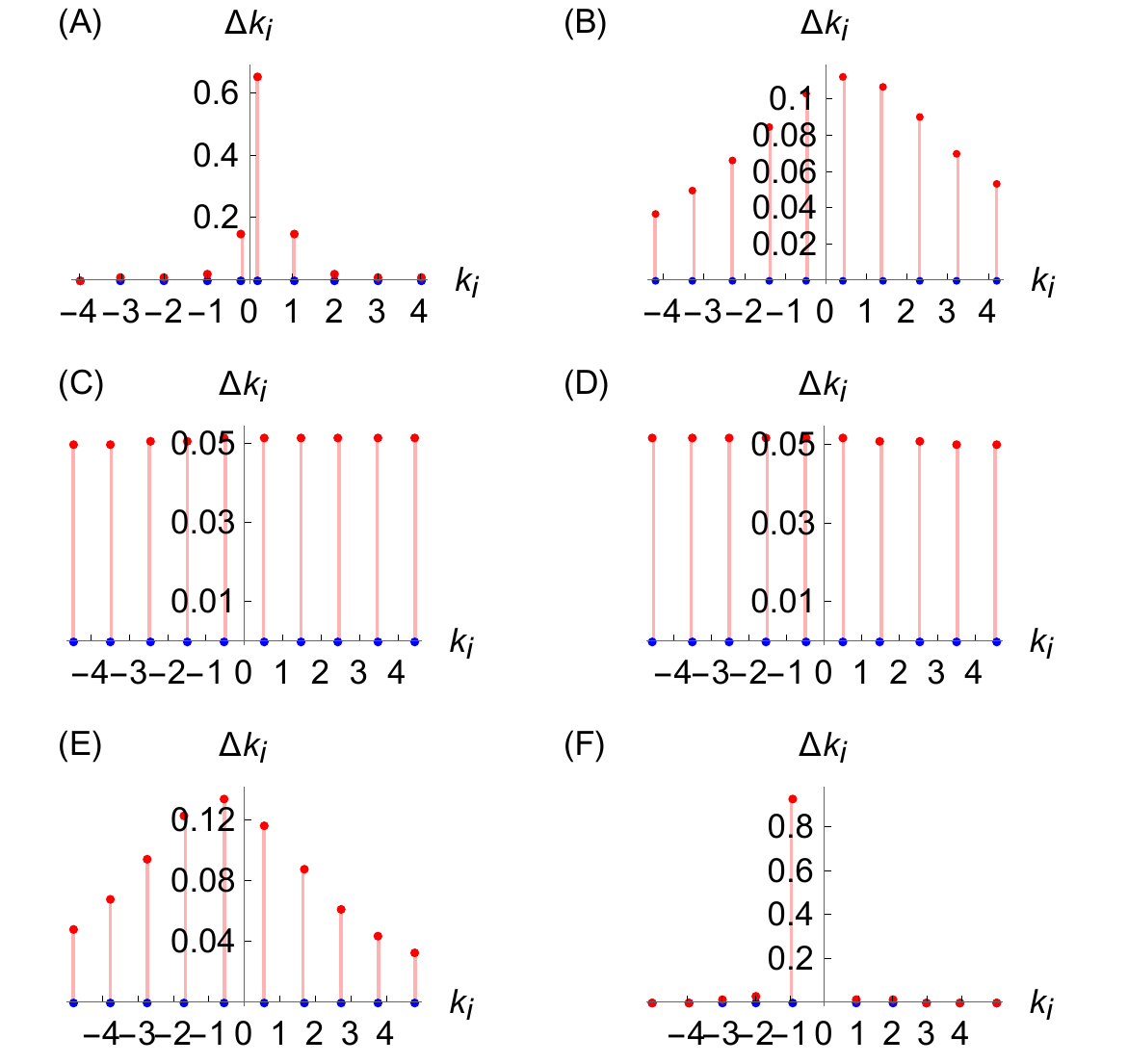}
\caption{(Color online.) Quasi-momentum($\{k_i\}$) distribution for the ground state (blue circle, x-axis) and the change($\{\Delta k_i\}$) after the lowest excitation (red circle, y-axis). The unit of momentum is $2\pi/L$. Here we have taken a few typical coupling strengths, and for (A-F) we have $(-k_F/(mg), E/E_F, m^*/m)=$$(-15\pi, 0.01, 1.)$(A), $(-0.5\pi, 0.3, 1.04)$(B), $(-0.05\pi, 0.82, 2.84)$(C), $(0.05\pi, 1.08, -3.72)$(D), $(0.5\pi, 1.66, -1.04)$(E), $(15\pi, 1.99, -1. )$(F). Here the total number of $N=20$. To see more clearly we have shown a few $k_i$ with lowest amplitudes.} \label{fig2}
\end{figure}

The evolution of $E$ and $m^*$ as shown  in Fig.\ref{fig1} can be traced back to the change of quasi-momentum ($\{k_i\}$) distribution as varying the couplings. In Fig.\ref{fig2}, 
we show the $\{k_i\}$ distribution for the ground state ($Q=0$) and the change of $\{k_i\}$ after the lowest excitation (to state with $Q=2\pi/L$), taking several typical values of $-1/g$ from (A) to (F). 

Fig.\ref{fig2} shows that at weak repulsion $g\rightarrow 0^+$ (A), $\{k_i\}$ for the ground state symmetrically distribute near the non-interacting limit: $k=0, \ \pm2\pi/L, \ \pm4\pi/L,...$(see blue circles), thus giving a small  $E$ (Fig.\ref{fig1}(a)). In this case the lowest excitation is dominated by a single excitation from $k\sim 0$ to $2\pi/L$ (see red circles), and such single-particle feature gives $m^*\sim m$ (Fig.\ref{fig1}(b)). As increasing $g$ (B), $\{k_i\}$ for the ground state deviate more from the non-interacting limit, and the excitation also expands to a broader range in $k$-space; 
accordingly $E$ and $m^*$ both increase. For large and positive $g$ (C), the ground state distribution is around $k=\pm\pi/L, \ \pm3\pi/L,...$, while during the excitation almost all $k_i$ are shifted equally in $k$-space with very small amplitude. Such collective behavior reflects the fermionized nature of a TG gas, and in the limit of $g\rightarrow +\infty$ this gives the polaron energy $E\rightarrow E_F$ and a huge 
$m^*\rightarrow Nm$, with $N$ the total number of the system. This means that during the excitation, all particles in the system move synchronously due to the strong correlations under hard-core interaction.

When go to the sTG side with a large and negative $g$(D), the ground state $\{k_i\}$ continue to expand in $k$-space, leading to a continuously increased $E$. The excitation also behaves collectively, while a subtle difference from (C) is that, here during the excitation the change of negative $k_i$ is larger than the change of positive ones, so to produce a large and negative $m^*$. This asymmetric excitation feature is more visible for an intermediate negative $g$ as shown in (E). In the limit of $g\rightarrow 0^-$ (F), the ground state $\{k_i\}$ distribute around $k=\pm2\pi/L, \ \pm4\pi/L,..$. Compare to the non-interacting limit, this corresponds to moving a pair of fermions from $k\sim 0$ to the Fermi surface $k\sim k_F$, which gives $E\rightarrow 2E_F$. 
In this case the excitation is dominated by a single change of $k$ from $-2\pi/L$ to $0$, leading to $m^*\sim -m$. 

To this end, we have shown that the behavior of $E$ and $m^*$ in Fig.\ref{fig1} can be well explained from the $\{k_i\}$ distributions shown in Fig.\ref{fig2}.

\subsection{Total spin} 

Now we come to the question whether the negative $m^*$ in $g<0$ side indicates instability towards the ferromagnetic state which lies below the repulsive branch (see Fig.\ref{fig1}(a)), or the phase separation which has ferromagnetic component. We address this question by investigating the total spin of repulsive Fermi polaron after the excitations. 
We will show in the following that all the excited states have the same total spin, $S=N/2-1$ (the lowest one for the polaron system), as the ground state, and there is no $S=N/2$ component. Therefore the negative $m^*$ and its associated excitations are 
irrelevant to the ferromagnetic instability. 

We first express the polaron wave function (\ref{wf}) in the coordinate(${\bf x}$) and spin(${\bf \xi}$) space:
\begin{equation}
\Psi({\bf x}, \xi)=\sum_{\alpha=1}^{N!}P_{\alpha}\Big( \psi({\bf x}) \xi_{\downarrow}(1)\xi_{\uparrow}(2)...\xi_{\uparrow}(N)\Big). \label{Psi}
\end{equation}
Here $P_{\alpha}$ is the permutation operator to guarantee the anti-symmetry of $\Psi$ with respect to  the simultaneous exchange of the coordinate and spin of any two particles. Given the expression of $\psi({\bf x})$ in (\ref{wf_2}), we can reorganize (\ref{Psi}) in terms of the particle order in coordinate space:
\begin{eqnarray}
\Psi({\bf x}, \xi)&=&\sum_{\alpha=1}^{N!}P_{\alpha}\Big\{ \theta(x_1<x_2<...<x_N)  \nonumber\\
&&\big[ \sum_{i=1}^N \phi_i({\bf x}) \xi_{\uparrow}(1)...\xi_{\downarrow}(i)...\xi_{\uparrow}(N) \big]\Big\}. \label{Psi_2}
\end{eqnarray}   
Here $\phi_1=\phi$ (Eq.\ref{wf_3}), while $\phi_i$($i=2,..N$) can be obtained from $\phi$ through the transformation: 
\begin{equation}
\phi_i=-\phi|_{x_{2,3..,i}\rightarrow x_{2,3..,i}+L; \ x_1\leftrightarrow x_i}. 
\end{equation}
It is straightforward to check that $\{\phi_i\}$ satisfy: 
\begin{equation}
\sum_i \phi_i ({\bf x})=0. \label{relation}
\end{equation}
Note that this relation only relies on the periodic boundary condition (\ref{PBC}). It will be important for the derivation of total spin as shown below. 

Now we study what $\Psi$(Eq.\ref{Psi_2}) produces when acted on the total spin operator ${\bf S}^2$. Since ${\bf S}^2$ does not modify the coordinate of fermions, it is adequate to only look at the wave function in the region $x_1<x_2<...<x_N$, i.e., within $[...]$ in (\ref{Psi_2}), which we denote as $\Phi$ from now on. For spin-1/2 fermions, we have ${\bf S}^2=(N^2-2N+4)/4+\sum_{\langle i,j \rangle} (s_i^+s_j^-+h.c.)$, and one can prove that the second part simply produces $-1$ after acting on $\Phi$ because of the relation (\ref{relation}).  Finally we can get:
\begin{equation}
{\bf S}^2 \Psi({\bf x},\xi)= \frac{N(N-2)}{4} \Psi({\bf x},\xi)
\end{equation}
i.e., $\Psi$ is an eigen-state of ${\bf S}^2$ with total spin $N/2-1$. 

Since above proof does not depend on the specific value of $c$ in Eq.\ref{eq_z},
the conclusion should apply to all states from exact solutions, including both the ground and excited states.  That is to say, the excitation of repulsive polaron will not change the total spin, and therefore the ferromagnetic state is irrelevant during the excitation. In fact, this can be understood by recalling that the ferromagnetic transition, as pointed out in Ref.\cite{Cui-Ho2}, refers to the level crossing between two orthogonal states with different total spins. Such transition cannot occur in realistic system unless one applies a tiny external field breaking the spin-rotation symmetry. Here since no symmetry-breaking field is applied during the polaron excitation, naturally the ferromagnetic state (or phase separation state with ferromagnetic component) should be excluded.  Therefore, the phenomenon of  negative $m^*$ can only reflect an intrinsic nature of the repulsive Fermi polaron itself.

\subsection{Pair correlation} 

It is insightful to see how the excitation changes the spin-spin correlation between the impurity ($\downarrow$) and majority fermions($\uparrow$). The pair-correlation function, which gives the relative probability of finding a $\downarrow$-spin at $x_1$ and a $\uparrow$-spin at $x_2$, can be written as\cite{McGuire1, McGuire2}:
\begin{equation}
\rho_{\downarrow\uparrow}(x_1,x_2)=\int dx_3...dx_N |\psi({\bf x})|^2,
\end{equation}
with $\psi({\bf x})$ expressed in (\ref{wf_2}). Due to the periodic boundary condition, here we only consider the region $x_1<x_2...<x_N$, i.e., the impurity is in the left side of all majority particles. Since $\rho_{\downarrow\uparrow}(x_1,x_2)$ only depends on the distance $x_2-x_1$, we can simply set $x_1=0$ and rewrite it as 
\begin{equation}
\rho(x)\equiv \rho_{\downarrow\uparrow}(0,x)=\int dx_3... dx_N |\phi(x_1=0, x_2=x, x_3...x_N)|^2
\end{equation}
with $\phi({\bf x})$ expressed in (\ref{wf_3}). $\rho(x)$ can be obtained following the mathematical tricks in Ref.\cite{McGuire2}. In order to capture the change of $\rho(x)$ during the excitation, we additionally carry out the normalization of $\rho(x)$ for both the ground state and the excited states, such that  $\int_0^L dx \rho(x)=N-1$. Finally we obtain 
\begin{equation}
\rho(x)=\Big( 2\sum_i \frac{|\alpha_i|^2}{L+\frac{|\alpha_i|^2}{mg}} \Big)^{-1} \sum_{i,j}\frac{|\alpha_i e^{ik_jx} - \alpha_j e^{ik_ix}|^2}{(L+\frac{|\alpha_i|^2}{mg})(L+\frac{|\alpha_j|^2}{mg})}. 
\end{equation}
Given the equality $\rho(x)=\rho(L-x)$, we only study the region $x\in[0,L/2]$.  

\begin{figure}[h]
\includegraphics[width=8cm, height=9.5cm]{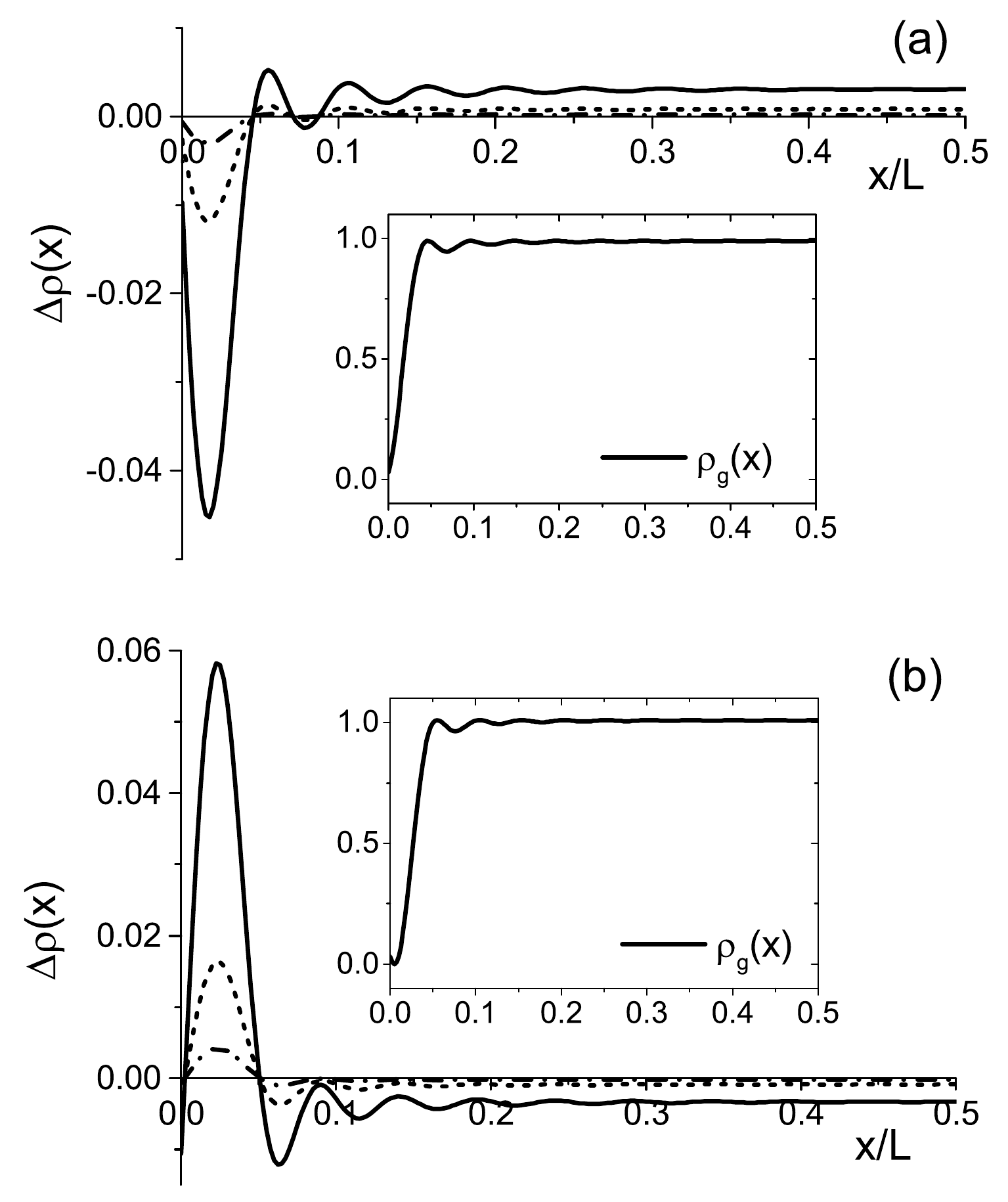}
\caption{Change of pair-correlation functions, $\Delta\rho(x)$, during the excitations to the 1st (dashed-dot), 2nd (short-dash) and 4th (solid) excited states. Here the $n$th excited state is with total momentum $Q=2n\pi/L$. (a) is for TG limit corresponding to (C) in Fig.\ref{fig2} with $m^*>0$; (b) is in sTG limit corresponding to (D) in Fig.\ref{fig2} with $m^*<0$. 
Insets of (a,b) show $\rho(x)$ for the ground state with $Q=0$. Here $N=20$.
} \label{fig3}
\end{figure}

In Fig.\ref{fig3} we show the change of $\rho(x)$ during the excitations, denoted as $\Delta\rho\equiv\rho_e-\rho_g$ with $\rho_{g(e)}$ the pair-correlation function for the ground (excited) state. Here we take two typical couplings in the TG and sTG regimes ((C) and (D) in Fig.\ref{fig2}), which are respectively associated with a large positive and a large negative $m^*$. In the insets of Fig.\ref{fig3} we show the ground state correlation denoted by $\rho_g(x)$. We see that $\rho_g$ behave quite similarly for the two couplings, which both show a dip near $x\sim 0$ because of the strong repulsion. Nevertheless, $\Delta\rho$ display remarkable difference between the two cases with opposite sign of $m^*$.
In the TG regime with a positive $m^*$ (Fig.\ref{fig3}(a)), $\Delta\rho$ is negative at small $x$ while positive at large $x$. This means that after the excitations, the probability of finding a majority fermion becomes smaller near the impurity while becomes larger far away from the impurity. In contrast, in the sTG regime with negative $m^*$ (Fig.\ref{fig3}(b)), $\Delta\rho$ displays a (positive) peak at small $x$ while becomes negative at large $x$. Intuitively, this means that during the excitation more majority fermions  move closer to the impurity instead of being repelled far away. Obviously this is contradictory to the instability towards phase separation as interpreted in Ref.\cite{Italy2}.

\section{Summary and discussion} 

In this work, we have utilized the exact solutions in 1D to reveal the ground state and excitation properties of repulsive Fermi polarons. In particular, we show that the negative effective mass of polarons in the sTG regime does not imply the instability to ferromagnetism or phase separation. Rather, it reflects an intrinsic excitation property of the repulsive polaron, as can be seen from the changes of quasi-momentum distribution and the pair correlation during the excitations. 

Our 1D results shed light on the observation of negative $m^*$ in the 3D Fermi polaron in recent LENS experiment\cite{Italy2}. These two systems share essential similarities in that, first, the negative $m^*$  both occur in the excited upper branch of two atomic systems with strong repulsion, and secondly, $m^*$  in both systems behave similarly as the repulsion energy increases, i.e., it undergoes a resonance structure from large repulsive to large negative. Given these similarities, understanding the negative $m^*$ in 1D will provide essential insights to the same phenomenon in 3D, although technically it is hard to describe the 3D system using the wave function and quasi-momentum language as in 1D case. 

In this context, an important contribution of the present work is that through the rigorous 1D analysis, this work raises another mechanism, instead of phase separation, for the negative $m^*$ of Fermi polarons and its associated instability in the strongly repulsive regime. In other words, the negative $m^*$ cannot be used as the determinant signature for phase separation. In addition, this work suggests that one cannot achieve phase separation from the low-lying excitations of a non-phase-separated system, unless there is a symmetry-breaking field. We hope our present work will stimulate more studies on the nature of repulsive Fermi polarons and their instabilities in high dimensions.

\section{Acknowledgment} 

We thank Nikolaj Zinner and Manuel Valiente for helpful discussions during the KITP program ``Universality in Few-Body Systems" in 2016. The work is supported by the National Key Research and Development Program of China (2016YFA0300603), and the National Natural Science Foundation of China (No.11622436, No.11374177, No.11421092, No.11534014).

\end{document}